\documentclass[conference,letterpaper]{IEEEtran}

\usepackage[utf8]{inputenc} 
\usepackage[T1]{fontenc}
\usepackage{url}
\usepackage{ifthen}
\usepackage{cite}
\usepackage[cmex10]{amsmath}
\usepackage{amssymb}
\usepackage{natbib}

\newtheorem{theorem}{Theorem}
\newtheorem{lemma}[theorem]{Lemma}

\begin{document}
\title{Super-Linear Growth of the Capacity-Achieving Input Support  
for the Amplitude-Constrained AWGN Channel} 

\author{%
  \IEEEauthorblockN{Haiyang Wang}
  \IEEEauthorblockA{Department of Applied and Computational Mathematics \\
                    Yale University \\
                    Email: haiyang.wang.hw578@yale.edu}
}

\maketitle

%%%%%%
%% Abstract: 
%% If your paper is eligible for the student paper award, please add
%% the comment "THIS PAPER IS ELIGIBLE FOR THE STUDENT PAPER
%% AWARD." as a first line in the abstract. 
%% For the final version of the accepted paper, please do not forget
%% to remove this comment!
%%

\begin{abstract}
% THIS PAPER IS ELIGIBLE FOR THE STUDENT PAPER AWARD. 
We study the growth of the support size of the capacity-achieving input distribution 
for the amplitude-constrained additive white Gaussian noise (AWGN) channel.
While it is known since \cite{smithInformationCapacityAmplitude1971} that the optimal input is discrete with finitely many mass points, 
tight bounds on the number of support points \(K_A\) as the amplitude constraint \(A\) increases remain open. 
Not much is known until recently, when \cite{dytsoCapacityAchievingDistribution2019} proved that $K_A$ grows at least linearly and at most quadratically in $A$. 
Here, we provide a novel method, 
building on~\cite{zhangDiscreteNoneinformativePriors1994,maBestApproximationFinite2024}, 
to derive the first non-trivial lower bound showing that \(K_A\) grows super-linearly in \(A\). 
\end{abstract}

\section{Introduction}

The amplitude-constrained AWGN channel
\begin{equation}
    Y = X + Z, \qquad Z \sim \mathcal{N}(0,1),
\end{equation}
is a fundamental model where the input \(X\) is restricted to \(|X|\le A\).
Its capacity is the maximal mutual information between the input and the output. 
\begin{equation}
    \label{eq:capacity}
    C(A) = \sup_{P_X:|X|\le A} I(X;Y).
\end{equation}
The input distribution that achieves the supremum in the RHS above, denoted by \(\pi_A^\star\), is called as the capacity-achieving input distribution. 
In the seminal work \cite{smithInformationCapacityAmplitude1971}, existence and uniqueness of capacity-achieving input distribution is established. Further more, it was proved that the \(\pi_A^\star\) is discrete with finitely many mass points \(K_A := |\operatorname{supp}(\pi_A^\star)|\), i.e., $\pi_A^\star = \sum_{n=1}^{K_A} p_n \delta_{x_n}$, for some $p_n \ge 0, \sum p_n = 1, x_n \in [-A,A]$. And, the capacity-achieving output distribution is a mixture of $K_A$ Gaussians with unit variance \(\phi * \pi_A^\star = \sum_{n=1}^{K_A} p_n \phi(x - x_n)\).
The main question is how does the discrete distribution $\pi_A^\star$ look like. And a particular question has been asked: how does $K_A$ scale with $A$? 

% This result, and later theoretical results on the capacity-achieving input distribution, relies much on the theory of complex analysis. 
\vspace{1em}

\noindent
\textbf{Related Work:}
\cite{zhangDiscreteNoneinformativePriors1994} provided many useful analysis and descriptions of the capacity-achieving input and output distributions --- 
in the bulk interior of $[-A,A]$, the output distribution is very close to the uniform distribution; 
near the boundary of $[-A,A]$, the output distribution deviates from the uniform distribution, and the bounds on the deviation are also established. 
Those analysis in \cite{zhangDiscreteNoneinformativePriors1994}, being highly technical and demanding lengthy analysis, are heavily leveraged in this paper. In addition, Zhang established that the $\pi_A^\star$, rescaled properly to $2A\pi_A^\star$, converges to certain improper priors as $A\to\infty$ in a weak sense: viewing the real line $\mathbb{R}$ as the limit $\lim_{A\to\infty} [-A,A]$, the $2A\pi_A^\star$ converges to the Jeffrey's prior on the real line; viewing the half real line $[0,\infty)$ as $\lim_{A\to\infty} [0,2A]$, and shift $2A\pi_A^\star$ to be supported on $[0,2A]$, then $2A\pi_A^*$ would converge to a improper prior consisting of discrete mass points along the half real line. The continuous improper prior on the whole of real line is essentially saying that in the bulk interior of $[-A,A]$, $\pi_A^\star$ is close to the uniform distribution in a weak sense. And the discrete improper prior on the half line is, on the other hand, characterizing how $\pi_A^\star$ behaves near the boundary of $[-A,A]$. 
A, understanding the improper prior on the half line is essential to answering the question of how $K_A$ scales with $A$. Lastly, Zhang conjectured that $K_A \propto A\log A$ as $A\to\infty$.

More recently, \cite{dytsoCapacityAchievingDistribution2019} proved the most tight upper and lower bounds on $K_A$ for the amplitude-constrained AWGN channel as follows: 
\[
\sqrt{1+ \frac{2A^2}{\pi e}} \le K_A \le a_2 A^2 + a_1 A + a_0,
\] 
where the upper bound is proven using zero-counting methods from complex analysis, and the lower bound using a simple entropy argument. 
The authors have conducted numerical experiments, based on which they conjectured that $K_A \propto A$.  

The larger scale and more universal numerical experiments in \cite{mattinglyMaximizingInformationLearned2018} suggested that $K_A \propto A^{4/3}$ as $A\to\infty$. Their numerical experiments is conducted on AWGN channels as well as a few more arbitrary channels that are quite different from AWGN channels: the binomial channel and some 2-dimensional channels. Across a variety of channels, their numerical results consistently suggested the scaling law $C \propto \frac34 \log K$, where $C$ is the capacity and $K$ is the number of mass points of the optimal input distribution. Despite the lack of a proof, the empirical universality across a variety of channels might be perceived as a strong evidence of their conjectured scaling law. 
% And, if such a scaling law is true universally across a variety of channels. 
 Furthermore, in a follow-up paper~\cite{abbottScalingLawDiscrete2019} the authors have provided a ``physicist's proof'' of this conjecture. 

Numerical experiments are susceptible to algorithmic bias and rounding errors, especially for this problem.  
Astonishingly, each research group that have conducted numerical experiments on this problem has concluded to a different scaling law of $K_A$ --- no two empirical scaling laws are the same~\cite{mattinglyMaximizingInformationLearned2018,
dytsoCapacityAchievingDistribution2019,
changCalculatingCapacityInfiniteInput1988}.  
And there is a principled reason for this discrepancy: the optimization problem and the algorithms inherently requires extreme high precisions. 
The rounding errors is particularly problematic for the amplitude-constrained Gaussian channels. For large $A$, the nearly-optimal output distribution in the bulk interior would be close to the uniform distribution, with differences at the scale of numerical precision. Moreover, if using the standard Blahut–Arimoto algorithm \cite{blahutComputationChannelCapacity1972,arimotoAlgorithmComputingCapacity1972}  to compute the optimal input distribution, as commonly done in the literature, it will be necessary to nested-ly apply numerical integrations on the log-probability of the output distribution, which introduces additional biases and errors. 
For the Blahut-Arimoto algorithm to work properly, it is required that the difference between the output distribution and the uniform distribution to be accurately reflected in the result of numerical integration. This requires not only a high precision of floating point arithmetic, perhaps beyond the standard double precision for only moderately large $A$, but also a well-designed quadrature rule for the numerical integration. 
All parts considered, the rounding errors, numerical integration biases, and the precision required for the Blahut-Arimoto algorithm to converge to the optimal input distribution, the numerical experiments are highly susceptible to incorrect conclusions. 
It might be of some interests to develop error analysis for these numerical experiments and the underlying algorithms, to gain more insights as to the reliability of the empirical results. 
We should be extremely cautious about the results from numerical experiments.

\vspace{1em}
\noindent
\textbf{Our Contributions:}
In this paper, we proved the first non-trivial lower bound on the growth rate of $K_A$, that is $K_A \gg A$ as $A\to\infty$. This rigorously disproves the conjecture that $K_A \propto A$ as $A\to\infty$ of~\cite{dytsoCapacityAchievingDistribution2019}.
\begin{theorem}
    \label{thm:main}
    As \(A \to \infty\), the support size of the optimal input grows super-linearly:
    \begin{equation}
        \lim_{A\to\infty} \frac{K_A}{A} = \infty.
    \end{equation}
\end{theorem}
Our method is quite novel and different in nature from the existing methods, which relies mostly on complex analytical ideas. On one hand, we rely on precise analysis of the behavior of the capacity-achieving output distribution developed in \cite{zhangDiscreteNoneinformativePriors1994}, concluding that the output distribution is very close to the uniform distribution: the $\chi^2$ divergence goes to 0 as $A\to\infty$. On the other hand, techniques from \cite{maBestApproximationFinite2024} are applied to develop a theory of best approximation by finite Gaussian mixtures: we are able to lower bound the $\chi^2$ divergence of any finite Gaussian mixture to the uniform distribution away from zero as long as $K \propto A$. Combining the two parts, it can be asserted that $K\gg A$, for otherwise would be a contradiction.  

\vspace{1em}
\noindent
\textbf{Organization:}
Section~\ref{sec:notation} introduces the preliminaries and defines some basic notations that will be used in this paper throughout. 
Section~\ref{sec:chi0} states the result on $\chi^2$-divergence of the capacity-achieving output distribution to the uniform distribution. 
Section~\ref{sec:chi-lb} presents a novel theory of best approximation by finite Gaussian mixtures in the $\chi^2$ divergence sense. 
With the results of these two sections, we are able to prove our main result Theorem~\ref{thm:main} in Section~\ref{sec:mainproof}. 
Section~\ref{sec:conclusion} concludes the paper. Proofs of the technical lemmas and theorems are provided in the appendix.

\section{Preliminary and Notation}
\label{sec:notation}

Let \(\phi(x) = \tfrac{1}{\sqrt{2\pi}} e^{-x^2/2}\) denote the standard Gaussian density.
For any input distribution \(\pi\), define the output density 
\(f_\pi := \phi * \pi\), where $*$ is the convolution operator of two probability densities. Therefore, given an random variable \(X\) with distribution $\pi$, the output random variable $Y = X + Z$ has density $f_\pi$. We define the following distributions that are of most interest in the following: the optimal output distribution for Equation~\eqref{eq:capacity} is $f_A^\star := \phi * \pi_A^\star$, and $f_{\text{unif,A}} := \phi * \operatorname{Unif}([-A,A])$, where $\operatorname{Unif}([-A,A])(x)$ is the density of the uniform distribution on $[-A,A]$. 

The $\chi^2$ divergence between two probability densities \(p,q\) on a measurable space $\mathcal{X}$ is defined as
\begin{equation}
    \chi^2(p\|q) = \int_{\mathcal{X}} \left(\frac{p(x)}{q(x)} - 1\right)^2 q(x) dx.
\end{equation}
For the ease of proofs and avoidance of some technicalities, we will not consider $\chi^2$ divergence over the real line. Instead, we will transform most of the probability distributions to the circle $[-\pi,\pi]$, and then compute the $\chi^2$ divergence over the circle. Here is how we do the transformation: given a probability density $p$, a random variable $X\sim p$, and the scale parameter $A$, we define the transformed random variable on the circle to be $[-\pi,\pi]\ni\tilde X = \frac{\pi}{A} (X \mod 2A)$, where $X\mod 2A:= X - 2A[ \frac{X}{2A} ] \in [-A,A]$ is the modulo operation. The transformed probability density is therefore $\tilde p(\tilde x) = \frac{A}{\pi}  \sum_{j\in\mathbb Z} p\left( \frac{A}{\pi} (\tilde x + 2\pi j) \right)$. It can be readily verified that $f_{\text{unif},A}=\phi * \operatorname{Unif}([-A,A])$ is transformed to $\tilde f_{\text{unif},A} = \operatorname{Unif}([-\pi,\pi])$. And $f_{A}^*$, the mixture of $K_A$ Gaussians of unit variance, is transformed to a mixture of $K_A$ wrapped Gaussians with variance $(\pi/A)^2$.

\section{Convergence of the Capacity-Achieving Output}
\label{sec:chi0}

We first show that the capacity-achieving output distribution approaches the uniform distribution, in the wrapped $\chi^2$ divergence sense, 
as the amplitude constraint $A$ grows to infinity.
\begin{theorem}
\label{thm:chi0}
As \(A\to\infty\),
\begin{equation}
\chi^2(\tilde f_A^\star, \operatorname{Unif}([-\pi,\pi])) \to 0.
\end{equation}
\end{theorem}
This result may be understood intuitively as follows. The capacity achieving output distribution $f_A^\star$ optimizes $I(X;Y) = H(Y) - H(Y|X) = H(Y) - \log(\sqrt{2\pi e})$; therefore, $f_A^\star$ is also the output distribution with maximum entropy. Moreover, when $A$ is large, $f_A^\star$ is mostly supported within the bulk interior of $[-A,A]$, with ignorable mass lying outside of the interval. For such a distribution to have maximum entropy, it should be quite close to the uniform distribution on $[-A,A]$, except near the boundary of $[-A,A]$. 
Therefore, the $\chi^2$ divergence between $\tilde f_A^\star$ and $\operatorname{Unif}([-\pi,\pi])$ should be small, with the only issue of boundary effects to be handled carefully. 
\cite{zhangDiscreteNoneinformativePriors1994} has provided very precise analysis on how close is $\tilde f_A^\star$ to the uniform distribution on the bulk interior of $[-A,A]$ and the bounds we need to control the boundary effects. His results are crucial to the proof of this theorem, which is provided in the appendix.

\section{Approximation Limits for Finite Gaussian Mixtures}
\label{sec:chi-lb}
Consider any input distribution \(\pi\) supported on \(K\) points,
and denote \(f_\pi = \phi * \pi\).
The next result quantifies how close such a mixture can be to the uniform output.
\begin{theorem}
\label{thm:chi-lb}
For every \(A>0\) and any discrete \(\pi\) with \(K\) mass points,
\begin{equation}
\chi^2(\tilde f_\pi, \operatorname{Unif}([-\pi,\pi])) \ge \frac12 \exp\left( - 4\pi^2\frac{K^2}{A^2} \right)
\end{equation}
\end{theorem}

This theorem is a variant of Theorem 6 and Theorem 7 of \cite{maBestApproximationFinite2024}: 
their Theorems lower bounds the $\operatorname{TV}$ instead of $\chi^2$, but fails to lower bound the divergence away from zero when $K=\Theta(A)$ as $A\to\infty$. Here, by focusing on the $\chi^2$ divergence and using similar steps from~\cite{maBestApproximationFinite2024}, we are able to lower bound the divergence away from zero whenever $K=O(A)$.

% ============================================================
\section{Proof of the Main Result}
\label{sec:mainproof}

Combining Theorems~\ref{thm:chi0} and~\ref{thm:chi-lb} yields the result of this paper.

\begin{IEEEproof}[Proof of Theorem~\ref{thm:main}]
Suppose by contradiction that $K_A = O(A)$. Then as $A \to \infty$, $K_A / A \le c$ for some constant $c$. Theorem~\ref{thm:chi-lb} gives 
\(\chi^2(\tilde f_A^\star, \operatorname{Unif}([-\pi,\pi])) \ge \frac12 \exp\left( - 4\pi^2 c^2 \right)\),
a fixed positive number. On the other hand, Theorem~\ref{thm:chi0} gives that 
\(\lim_{A\to\infty} \chi^2(\tilde f_A^\star, \operatorname{Unif}([-\pi,\pi])) \to 0\). 
This is a contradiction, therefore we have that $K_A \gg A$ as $A \to \infty$.
\end{IEEEproof}

\section{Conclusion}
\label{sec:conclusion}

We have proved that the support size $K_A$ of the capacity-achieving input distribution for the amplitude-constrained AWGN channel 
grows super-linearly with the amplitude $A$.
This is the first non-trivial lower bound on $K_A$, and rigorously disproves the standing conjecture that $K_A = \Theta(A)$ from \cite{dytsoCapacityAchievingDistribution2019}. 
Our method is quite novel in this problem. It firstly made the simple observation that the capacity-achieving output distribution is roughly uniform. Then, it developed a novel theory of best approximation by finite Gaussian mixtures. By combining the two, it is shown that it is required for the optimal input to have a super-linearly growing $K_A$ for it to achieve the capacity.

This work is only a first step using the new theory. The quantitative description of capacity-achieving output distribution from \cite{zhangDiscreteNoneinformativePriors1994} might be improved to yield a better estimation of the growth rate of $K_A$. This will be left for future work.

\section{Acknowledgment}
\label{sec:acknowledgment}
The author is deeply grateful to Professor Yihong Wu for bringing this problem to the author's attention and for numerous valuable discussions and suggestions throughout the course of this research.

\cite{*}

\bibliographystyle{IEEEtran}
% \bibliographystyle{plainnat}
% \bibliography{capacity_references}
\bibliography{main2.bbl}

\appendix

In this appendix, we provide proofs for Theorems~\ref{thm:chi0} and~\ref{thm:chi-lb}.

\subsection{Proof of Theorem~\ref{thm:chi0}}
Before proving Theorem~\ref{thm:chi0}, we describe the following lemmas, mostly due to~\cite{zhangDiscreteNoneinformativePriors1994}.
\begin{lemma}
\label{lem:key}
For $0<B<A$, as long as $A \to \infty$ and  $A-B \to \infty$, we have that
\begin{align}
    \lim_{A \to \infty, A-B \to \infty} \sup_{|x| \le B} |2Af_A^\star(x) - 1| = 0
    \label{eq:key}
\end{align}
\end{lemma}
This lemma appears as Theorem 4.16 in~\cite{zhangDiscreteNoneinformativePriors1994}. It means that in the bulk interior of $[-A,A]$, the output distribution is very close to the uniform distribution.

\begin{lemma} \label{lem:trivial-bound}
Given a mixture of Guassians $f = \phi * \pi$ with $\pi$ supported in $[-A,A]$, with $A>1$,  we have the following bound on $\tilde f$
\begin{align}
    \sup_{\tilde x} \tilde f(\tilde x) 
    & \le \frac{A}{\pi}\left(3 \sup_x f(x) + \frac{1}{\exp(A/2)-1}\right) 
\end{align}
\end{lemma}

\begin{IEEEproof}
\begin{align}
\tilde f(\tilde x) 
&= \frac{A}{\pi} \sum_{j\in\mathbb Z} f(x + 2Aj) \\
& \le \frac{A}{\pi} \left(3 \bar f + \sum_{|j| > 1} f(x + 2Aj)\right) \\ 
& < \frac{A}{\pi} \left(3 \bar f + 2\sum_{j >= 0} \phi((2j+1)A)\right) 
\end{align}
where $x = \frac{A}{\pi}\tilde{x}\in[-A,A]$, $\bar f = \sup_x f(x)$, and the $2\sum_{j >= 0} \phi((2j+1)A)$ term can be easily bounded by a geometric series. 

\end{IEEEproof}

Now, it is necessary to introduce several notions described in Chapter 4, Section 5 of \cite{zhangDiscreteNoneinformativePriors1994} in order to handle the boundary effects. The shifted and rescaled density, $\pi_A'(x):= 2A \pi_A^\star(x-A)$, is a discrete density supported on $[0,2A]$. In the limit of $A\to\infty$, $\pi_A'(x)$ converges to a improper prior $\pi'$. And there are three relevant improper output densities that is of usage later: $f_A' = \phi * \pi_A', \, g = \phi * \pi'$, which can be alternatively defined as $g(x) = \lim_{A\to\infty} f_A'(x)$, and lastly $g_A = \phi * (\pi'_A \operatorname{1}_{[0,A]} +  \operatorname{1}_{[A,\infty]})$. And there are a few important inequalities among these quantities that will be used later. 
\begin{lemma}
    \label{lem:improper-bounds}
    We have the following bounds: 
    \begin{align}
        \label{eq:aaa}
        \lim_{A\to\infty} \sup_{x\le A} |f_A'(x) - g_A(x)| &\to 0 \\
        \label{eq:bbb}
        \lim_{A\to\infty} \sup_{x\le A} |g(x) - g_A(x)| &\to 0 \\
        \label{eq:ccc}
        \sup_x g(x) &< \infty 
    \end{align}
\end{lemma}
All three equations are proved in \cite{zhangDiscreteNoneinformativePriors1994}. More specifically, 
\eqref{eq:aaa} is from page~79, \eqref{eq:bbb} is by Theorem 4.19, and \eqref{eq:ccc} is a simple consequence of equation (4.51).

With these lemmas, we can prove Theorem~\ref{thm:chi0} as follows.
\begin{IEEEproof}[Proof of Theorem~\ref{thm:chi0}]
The concerned quantity is 
\begin{align}
\chi^2(\tilde f_A^\star, \operatorname{Unif}([-\pi,\pi])) = 2\pi \int_{-\pi}^{\pi} \left(\tilde f_A^\star(\tilde x) - \frac{1}{2\pi}\right)^2 d\tilde x 
\label{eq:chi2-integral}
\end{align}

Therefore, we need to control the integrand $\tilde f_A^\star(\tilde x) - \frac{1}{2\pi}$. We break the integral into three intervals and control the integrand in each interval separately: $[-\pi, -\pi+\delta], [-\pi+\delta, \pi-\delta], $ and $[\pi-\delta, \pi]$, with $\delta = \frac{\pi}{\sqrt{A}}$. We call the first and third intervals the boundary intervals, and the middle interval as the bulk interior interval. 

In the bulk interior interval $\tilde x\in[-\pi+\delta, \pi-\delta]$, we have that
\begin{align}
    \tilde f_A^\star(\tilde x) 
    &= \frac{A}{\pi} \sum_{j\in\mathbb Z} f_A^\star(\frac{A}{\pi}(\tilde x + 2\pi j)) \\
    &= \frac{A}{\pi} f_A^\star(x) + \frac{A}{\pi} \sum_{j\in\mathbb Z,j\ne0} f_A^\star(x+2Aj)
\end{align}
where $x = \frac{A}{\pi}\tilde x \in [-A+\sqrt{A},A-\sqrt{A}]$. Therefore, we have that 
\begin{align}
    |\tilde f_A^\star(\tilde x) - \frac{1}{2\pi}| \le \frac{1}{2\pi} |2A f_A^\star(x) - 1| + \frac{A}{\pi}| \sum_{j\ne0} f_A^\star(x+2Aj)|
\end{align}
The first term uniformly goes to 0 in the bulk interior interval as $A\to\infty$ by Lemma~\ref{lem:key}. The second term also uniformly goes to 0 as by the following elementary estimate: 
\begin{align}
    |\sum_{j\ne0} f_A^\star(x+2Aj)| 
    &\le 2\sum_{j\ge0} \phi(\sqrt{A}+2Aj) \\
    &< \sum_{j\ge0} \exp\left( -\frac{1}{2}(A+4Aj)\right)\\
    &= \frac{\exp(-\frac12A)}{1-\exp(-2A)}
\end{align}
Over the bulk interior interval $[-\pi+\delta, \pi-\delta]$, the integrand uniformly goes to 0 as $A\to\infty$, therefore the integral \eqref{eq:chi2-integral} over the bulk interior interval goes to 0 as $A\to\infty$. 

Now, we need to handle the integral \eqref{eq:chi2-integral} over the boundary intervals to show that the integral also goes to 0 as $A\to\infty$. With out loss of generality, we will focus on the left boundary interval $[-\pi, -\pi+\delta]$, as the same bounds hold for the right boundary interval $[\pi-\delta, \pi]$. In the following, we will show that the integrand is bounded by a constant; and then, since the integral is over an interval of diminishing length $\delta=\frac{\pi}{\sqrt{A}}$, the integral goes to 0 as $A\to\infty$. 

To bound the integrand $\left(\tilde f_A^\star(\tilde x) - \frac{1}{2\pi}\right)^2$ over the left boundary interval $[-\pi, -\pi+\delta]$, we only need to bound $\sup_{\tilde x}\tilde f_A^\star(\tilde x)$ by a constant. By the logic of Lemma~\ref{lem:trivial-bound}, we only need to prove that: 
\begin{align}
    \sup_{x\in[-A,-A+\sqrt{A}]} 2Af^\star_A(x) = \sup_{x\in[0,\sqrt{A}]} f'_A(x) = O(1)
\end{align}
This is a straightforward consequence of Lemma~\ref{lem:improper-bounds}.
\begin{align}
    f'_A(x) = |f'_A(x) - g_A(x)| + |g_A(x) - g(x)| + |g(x)|
\end{align}
Taking $\sup$ in $x\in[0,\sqrt{A}]$, and letting $A\to\infty$, we can apply \eqref{eq:aaa},\eqref{eq:bbb}, and \eqref{eq:ccc} to get the desired bound.  

Now, we have proved that the integral \eqref{eq:chi2-integral}, over the bulk interior, and the boundary intervals, goes to 0 as $A\to\infty$. Therefore, we have that
\begin{align}
    \chi^2(\tilde f_A^\star, \operatorname{Unif}([-\pi,\pi])) \to 0 \text{ as } A\to\infty
\end{align}

\end{IEEEproof}

\subsection{Proof of Theorem~\ref{thm:chi-lb}}
Before we proceed to prove Theorem~\ref{thm:chi-lb}, let us first introduce some preliminary notation. The $n$-th trigonometric moment of a random variable $X$ supported on the circle $[-\pi,\pi]$ is defined as $t_n(X) = \mathbb E [\exp(i n X)]$. Furthermore, we can define the $n$-th trigonometric moment matrix as follows:
$$
T_n(X) = \begin{pmatrix}
t_0(X) & t_1(X) & \cdots & t_{n}(X) \\
t_{-1}(X) & t_0(X) & \cdots & t_{n-1}(X) \\
\vdots & \vdots & \ddots & \vdots \\
t_{-n}(X) & t_{-n+1}(X) & \cdots & t_{0}(X)
\end{pmatrix} 
$$
which is an $(n+1) \times (n+1)$ Hermitian matrix. 

Now, given a discrete distribution $\pi = \sum_{k=1}^K p_k \delta_{x_k}$ with $K$ mass points, and a r.v. $X\sim \pi$, we can see that 
$T_n(X) = \sum_{k=1}^K p_k T_n(x_k)$ is sum of $K$ rank-1 matrices, therefore it has rank at most $K$. 

Now, we can begin to prove Theorem~\ref{thm:chi-lb}. This proof adapts the trigonometric moment method of Theorem 6 in \cite{maBestApproximationFinite2024}. 
\begin{IEEEproof}[Proof of Theorem~\ref{thm:chi-lb}]
The concerned distribution $\tilde f_{\pi}$ is a mixture of $K$ wrapped Gaussians of variance $\sigma^2=\frac{\pi^2}{A^2}$ on the circle $[-\pi,\pi]$. Let there be a  random variable $\tilde X\sim\tilde \pi$. Then, we have that: 
\begin{align}
    \chi^2(\tilde f_\pi, \operatorname{Unif}([-\pi,\pi])) 
    &= 2\pi \int_{-\pi}^{\pi} \left( \tilde f_\pi(\tilde x) - \frac{1}{2\pi} \right)^2 d\tilde x \\
    &= \sum_{n\ne 0} \exp(-\sigma^2 n^2) |t_n(\tilde X)|^2 
\end{align}
The above is consequences of the Parseval's theorem. Further more, we have that the $\chi^2$ divergence can be lower bounded as follows: 
\begin{align*}
    &> \sum_{n\ne0, |n| \le 2K} \exp(-\sigma^2 n^2) |t_n(\tilde X)|^2   \\
    &> \exp(-\sigma^2 4K^2) \sum_{n\ne0, |n| \le 2K}  |t_n(\tilde X)|^2  \\
    &> \exp(-\sigma^2 4K^2) \sum_{n\ne0, |n| \le 2K}  \frac{2K+1-n}{2K+1} |t_n(\tilde X)|^2  \\
    &= \frac{\exp(-\sigma^2 4K^2)}{2K+1} \sum_{n\ne0, |n| \le 2K} (2K+1-n) |t_n(\tilde X)|^2  \\ \\
    &= \frac{\exp(-\sigma^2 4K^2)}{2K+1} \|T_{2K}(\tilde X) - I\|_{\text{fro}}^2
\end{align*}
where $\| \cdot \|_{\text{fro}}$ is the Frobenius norm of a matrix. $T_{2K}(\tilde X)$ is a rank-$K$ matrix, therefore, by Eckart-Young-Mirsky theorem, we have that $\|T_{2K}(\tilde X) - I\|_{\text{fro}}^2 \ge K+1$. Substituting this into the above inequality, we obtain:
\begin{align*}
    > \frac{\exp(-\sigma^2 4K^2)(K+1)}{2K+1}   > \frac{1}{2}\exp(-4 \pi^2 \frac{K^2}{A^2})
\end{align*}

\end{IEEEproof}

\end{document}